\documentclass[aps,preprint,amssymb,superscriptaddress]{revtex4-1}
\usepackage{floatrow}
\newfloatcommand{capbtabbox}{table}[\FBwidth]
\usepackage{graphicx}% Include figure files
\usepackage{dcolumn}% Align table columns on decimal point
\usepackage{bm}% bold math
\usepackage{natbib}
\usepackage{CJK}
\usepackage[utf8]{inputenc}
\usepackage[T1]{fontenc}
\usepackage{mathptmx}
\usepackage{gensymb}
\usepackage{graphicx}% Include figure files
\usepackage{epstopdf}
\usepackage{dcolumn}% Align table columns on decimal point
\usepackage{bm}% bold math
\usepackage{amsmath}
\usepackage{mathtools}
\usepackage{relsize}
\usepackage{multirow}
\usepackage[colorlinks]{hyperref}
\begin{document}

%\preprint{AIP/123-QED}

%\begin{CJK*}{GB}{}
%\preprint{APS/123-QED}
\title{Theoretical Investigation of Charge Transfer Between Two Defects in a Wide-Bandgap Semiconductor}% Force line breaks with \\

\author{Rodrick Kuate Defo \footnote{E-mail: rkuatedefo@princeton.edu}}
\affiliation{Department of Electrical and Computer Engineering, Princeton University, Princeton, NJ 08540}
\author{Alejandro W. Rodriguez}
\affiliation{Department of Electrical and Computer Engineering, Princeton University, Princeton, NJ 08540}
\author{Efthimios Kaxiras} 
\affiliation{John A. Paulson School of Engineering and Applied Sciences, Harvard University, Cambridge, MA 02138, USA}
\affiliation{Department of Physics, Harvard University, Cambridge, MA 02138, USA}
\author{Steven L. Richardson} 
\affiliation{John A. Paulson School of Engineering and Applied Sciences, Harvard University, Cambridge, MA 02138, USA}
\affiliation{Department of Electrical and Computer Engineering, Howard University, Washington, DC 20059}
\date{\today}% It is always \today, today,

\begin{abstract}

Charge traps in the semiconductor bulk (bulk charge traps) make it difficult to predict the electric field within wide-bandgap semiconductors. The issue is the daunting number of bulk charge-trap candidates which means the treatment of bulk charge traps is generally qualitative or uses generalized models that do not consider the trap's particular electronic structure. The electric field within a wide-bandgap semiconductor is nonetheless a crucial quantity in determining the operation of semiconductor devices and the performance of solid-state single-photon emitters embedded within the semiconductor devices. In this work we accurately compute the average electric field measured at the location of N$V^-$ charged defects for the substitutional N (N$_\text{C}$) concentration of $n_{\text{N}_\text{C}} \approx 1.41\times10^{18}$~cm$^{-3}$ for the commonly used oxygen-terminated diamond (see [D. A. Broadway \textit{et al}., Nature Electronics 1, 502 (2018)]). We achieve this result by evaluating the leading-order contribution to the electric field far away from the surface, which comes from the N$_\text{C}$ defects that induce the ionization of the N$V^-$. Our results use density-functional theory (DFT) and the principle of band bending. Our work has the potential to aid both in the prediction of the functioning of semiconductor devices and in the prediction and correction of the spectral diffusion that often plagues the optical frequencies of solid-state single-photon emitters upon repeated photoexcitation measurements. Our results for the timescales involved in thermally driven charge transfer also have the potential to aid in investigations of charge dynamics. 
%We show in this work that when X$V^-$ color centers exhibit a has been shown to mediate entanglement between electronic spins and surrounding nuclear spins. This entanglement leads to proximal nuclear spins enhancing electronic spin coherence and distant nuclear spins destroying it.   This result suggests that if samples are cooled to a low enough temperature that a Jahn-Teller distortion emerges, if such a distortion does not favor enhanced spin coherence, reheating and repeating the process may ultimately coax the system into a distortion that does. Here the enhancement in spin coherence would be the result of the modulation of the hyperfine interaction strength between the X$V^-$ center electronic spin and the neighboring $^{13}$C nuclei when the center is cycled between the three energetically equivalent distortions. 
\end{abstract}

\maketitle
%\preprint{APS/123-QED}

\section{INTRODUCTION}
The ability to set the electric field within a semiconductor device to very precise values is essential for the functioning of the device~\cite{Dolde2011electric,Broadway2018spat,Iwasaki2017direct,Zhang2012band,Kotadiya2018universal,Simon2010polarization,Stathis2006the,Zhang2006electronic,Kaczer2018a}. This ability is impeded by inhomogeneities in the semiconductor device, such as those due to charge traps in the semiconductor bulk (bulk charge traps)~\cite{Broadway2018spat}, which can lead to failure of the semiconductor device~\cite{Iwasaki2017direct}. In the context of next-generation semiconductor devices incorporating point-defect qubit candidates, bulk charge traps can cause spectral diffusion of the optical frequencies of the point-defect qubit candidates upon repeated photoexcitation measurements~\cite{Bassett2011electrical,Forneris2018mapping,McCullian2022quantifying}, limiting the ability to achieve long-distance entanglement of photons for which indistinguishability of the photons is needed~\cite{Machielse2019quantum}. Due to the significant number of potential bulk charge-trap candidates, the treatment of the effect of bulk charge traps on the electric field within semiconductor devices has largely been qualitative or has employed generalized models that do not consider the trap's particular electronic structure~\cite{Pierre2009background}. A consequence of the difficulty associated with accurately simulating the electric field within semiconductor devices has been interest in monitoring of the electric field \textit{in situ}, using for example the optical frequencies of N$V^-$ single-photon emitters in diamond~\cite{Dolde2011electric,Iwasaki2017direct,Broadway2018spat,Forneris2018mapping}. Given that the N$V^-$ is being used as an \textit{in situ} sensor to measure a critically important quantity in the wide-bandgap semiconductor that is diamond and that bulk charge traps demonstrably affect the optical frequencies through which this measurement is performed, it is necessary to gain some theoretical insight into the exact extent to which the measured field is influenced by bulk charge traps.

Based on density-functional theory (DFT)~\cite{Kaxiras2003atomic,Kresse1993ab,Kresse1996efficient,Kresse1999from,Heyd,Krukau} and the principle of band bending~\cite{Zhang2012band,Broadway2018spat}, we elucidate the experimental measurement using optically detected magnetic resonance (ODMR) spectroscopy of N$V$ centers of an average electric field of 291$\pm5$~kV~cm$^{-1}$ for the substitutional N (N$_\text{C}$) concentration of $n_{\text{N}_\text{C}} \approx 1.41\times10^{18}$~cm$^{-3}$ for the commonly used oxygen-terminated diamond~\cite{Broadway2018spat}. Our results could ultimately help predict the functioning of semiconductor devices as rectifiers and switching devices, where built-in defect-induced fields would lead to losses. Additionally, our results could be useful in predicting and in correcting the spectral diffusion of the optical frequencies of solid-state single-photon emitters as relates to their use for applications in quantum information and computation.

This work is organized as follows. The computational tools used in this work will be presented in Section \ref{sec:methodology}. Next, Section \ref{sec:methods} will provide details regarding the theoretical formalism employed in this work. We will devote Section \ref{sec:imp} to a discussion of the Broadway \textit{et al.} experiment~\cite{Broadway2018spat} and to our elucidation of their experimental measurement of an average electric field of 291$\pm5$~kV~cm$^{-1}$ for the N$_\text{C}$ concentration of $n_{\text{N}_\text{C}} \approx 1.41\times10^{18}$~cm$^{-3}$ due to band bending for the commonly used oxygen-terminated diamond~\cite{Broadway2018spat}. Finally, our conclusions will be presented in Section \ref{sec:conc}.

\section{Computational methods \label{sec:methodology}}
 Our bandstructure calculations used VASP~\cite{Kresse1993ab,Kresse1996efficient,Kresse1999from} with the screened hybrid functional of Heyd, Scuseria and Ernzerhof (HSE06)~\cite{Heyd,Krukau}. We performed atomic-position relaxations for the primitive fcc unit cell of diamond, which were terminated when the forces dropped below a threshold of $10^{-2}$ eV$\cdot$\AA$^{-1}$. The wavefunctions for the primitive unit cell were expanded in a planewave basis with a cutoff energy of 500~eV. The primitive unit cell of diamond contains 2 atoms and was relaxed using a $\Gamma$-centered grid of $7\times7\times7$ k-points. Bandstructures were calculated for the primitive unit cell applying spin polarization with 20, 50, 100, and 200 reciprocal lattice points along the lines connecting consecutive high-symmetry points in the bandstructure path.
 
 Our formation energy calculations also used VASP with the screened hybrid functional of Heyd, Scuseria and Ernzerhof (HSE06). Our calculations were terminated when the forces in the atomic-position relaxations dropped below a threshold of $10^{-2}$ eV$\cdot$\AA$^{-1}$. The wavefunctions were expanded in a planewave basis with a cutoff energy of 430~eV, the size of the supercell was 512 atoms ($4\times4\times4$ multiple of the conventional unit cell), and $\Gamma$-point integration was used. The elements used in our calculations and the associated ground-state structures and values of their chemical potentials are: N ($\beta$ hexagonal close-packed structure, $-11.39$ eV/atom) and C (diamond structure, $-11.28$ eV/atom). We note that all of the formation energies used in this work were computed for defects in their ground state.

\section{THEORETICAL APPROACH AND DISCUSSION \label{sec:methods}}

\subsection{Overview of the Investigated Species and of the Thermodynamically Relevant Quantities}
\begin{figure}[ht!] 
\centering
\includegraphics[width=0.75\textwidth]{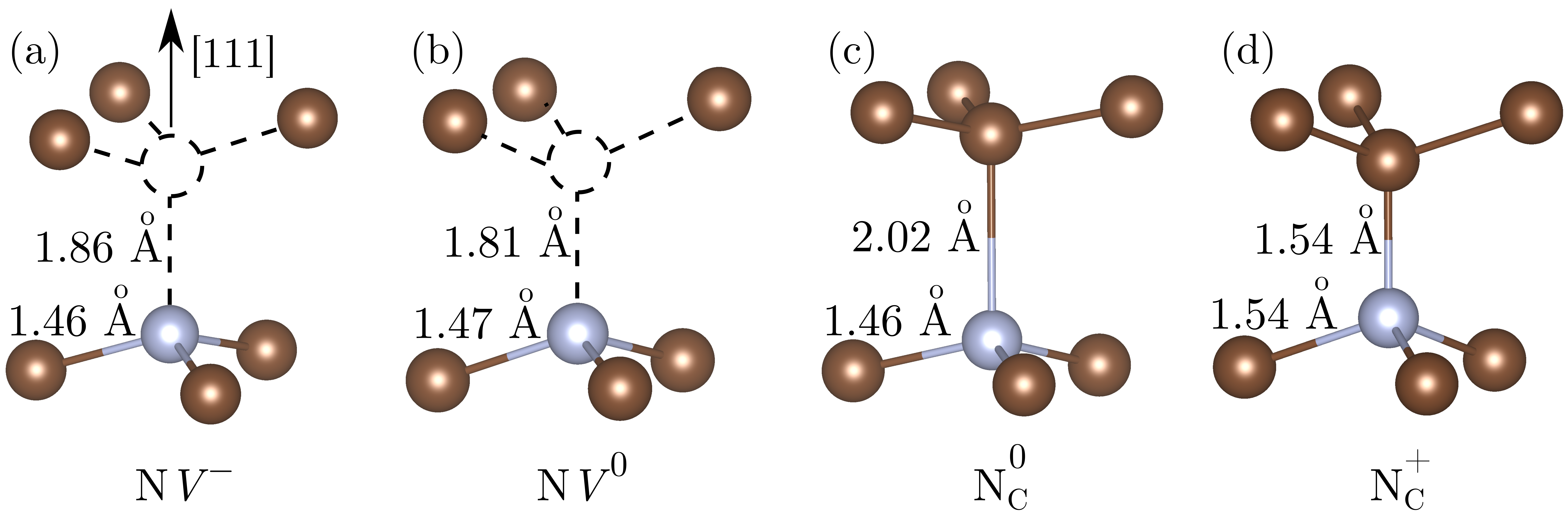}
\caption{Structure of (a) N$V^-$, (b) N$V^0$, (c) N$_\text{C}^0$, and (d) N$_\text{C}^+$. Carbon atoms are shown in brown and nitrogen atoms in purple. Carbon vacancies are shown as dashed circles. The distance between the C vacancy ($V$) and the N atom in the N$V$ defects was obtained as described in previous work~\cite{Kuate2021theor}. All defects have the orientation indicated in (a).} 
\label{fig:structures}
\end{figure}

Before proceeding to motivate our approach, we will briefly describe the defect species investigated in this work and define the thermodynamic quantities that are relevant for our investigation. The defect species investigated in this work included the nitrogen-vacancy center (N$V$) in diamond in the singly negatively charged (N$V^-$) and neutral (N$V^0$) charge states. In diamond, the nitrogen-vacancy center consists of a single N atom adjacent to a single C vacancy. In both the negative and neutral charge states, the defect has $C_{3v}$ symmetry due to breaking of tetrahedral symmetry as a result of the presence of the single C vacancy adjacent to the N atom. The substitutional N defect (N$_\text{C}$) was also investigated, which consists of a single N atom in a C position in diamond. As found in previous theoretical work~\cite{Kuate2021theor,Mainwood,Kajihara,Jones,Briddon1992,Lombardi_2003} and experimental work~\cite{Cook1966,Ammerlaan1981electron,Smith1959}, the neutral state (N$_\text{C}^0$) exhibits $C_{3v}$ symmetry due to elongation of a single N-C bond relative to the other three N-C bonds, while the singly positively charged state (N$_\text{C}^+$) exhibits tetrahedral symmetry. The structures of the defects are depicted in Fig. \ref{fig:structures}.  

We now turn to a discussion of the thermodynamic quantities that are relevant for our investigation. The equilibrium concentration of a charged defect species, $n_{\text{X}^\text{q}}$, in a semiconductor crystal is given by~\cite{Buckeridge2019equilibrium}

\begin{equation}
\label{eq:concentration}
    n_{\text{X}^\text{q}} = N_{\text{X}}g_{\text{X}^{\text{q}}}\exp(-\Delta H_f({\rm X^{\rm q}}, \, \{\mu_i^\text{X}\}, \, E_\text{F})/k_BT),
\end{equation}
where X is the defect species, $\text{q}$ is the charge of X, $N_\text{X}$ is the concentration of crystal sites on which X can form, $g_{\text{X}^{\text{q}}}$ is the degeneracy arising from the symmetry of $\text{X}^\text{q}$, $\{\mu_i^\text{X}\}$ denotes the set of chemical potentials for the constituent atoms of the defect $\text{X}$, $E_{\text{F}}$ is Fermi level, $k_B$ is Boltzmann's constant, and $T$ denotes temperature. The formation energy of X$^{\rm q}$ is $\Delta H_f({\rm X^{\rm q}}\, \{\mu_i^\text{X}\}, \, E_\text{F})$, which is given by~\cite{zhang1991chemical, Freysoldt2014first,Kuate2018energetics,Kuate2019how,Kuate2021methods,Kuate2021theor,zunger2021under,Yang2015self,Ashcroft1976solid}

\begin{equation}
\label{eq:form_eq-1.0}
\Delta H_f({\rm X^{\rm q}}, \, \{\mu_i^\text{X}\}, \, E_\text{F}) = E_{\text{def}}({\rm X^{\rm q}}) - E_0 - \sum_i\mu_i^\text{X}n_i + {\rm q}\, E_{\text{F}} + E_{\text{corr}}(\rm X^{\rm q}),
\end{equation}
where $E_{\text{def}}({\rm X^{\rm q}})$ is the energy of the charged supercell with the X$^{\rm q}$ species, $E_0$ is the energy of the stoichiometric neutral supercell, $\mu_i^\text{X}$ is the chemical potential of the $i^{\rm th}$ atomic constituent that was removed from or added to the stoichiometric supercell to produce the supercell with X (as above, $\{\mu_i^\text{X}\}$ denotes the set of all such chemical potentials), $n_i$ is a positive (negative) integer representing the number of the $i^{\rm th}$ constituent that was added (removed) to produce X, $E_{\text{F}}$ is again the Fermi level and is treated as a parameter, and $E_{\text{corr}}({\rm q})$ is an electrostatic correction term. 

The importance of the 
term $E_{\text{corr}}({\rm q})$, which is introduced as a correction to account for a finite supercell when performing a calculation for a charged defect, and the method for calculating it has been outlined by previous authors~\cite{Vinichenko,Freysoldt2011electrostatic,Freysoldt2009fully,Kumagai,Komsa2013finite,Walsh2021}.
To briefly motivate the importance of the calculation of $E_{\text{corr}}({\rm q})$, the use of a periodic supercell to calculate the total energy of a charged defect naturally leads to divergence of the total energy due to infinitely many uncompensated charges. In order to remedy the issue, a neutralizing background charge is applied which causes spurious terms to arise in the total energy~\cite{Vinichenko,Castleton2006managing,Komsa2012comparison,Alkauskas,Freysoldt2009fully,Freysoldt2011electrostatic,Komsa2012finite,Kumagai,Castleton2009density}. The energy, $E_{\text{corr}}({\rm q})$, is designed precisely to correct these spurious terms.

The formation energy $\Delta H_f({\rm X^{\rm q}}, \, \{\mu_i^\text{X}\}, \, E_\text{F})$ depends on the parameterized Fermi level, $E_\text{F}$, and for every charge state there will be a single range of $E_\text{F}$ where that charge state has the minimum formation energy. This fact allows us to define an adiabatic charge-transition level (ACTL)  $\epsilon^{\rm X}(\text{q}/\text{q}^{\prime})$ which is the value of the Fermi level $E_{\text{F}}^*$ for which X$^\text{q}$ and $\text{X}^{\text{q}^{\prime}}$ have equal formation energies, 
\begin{equation}
\label{eq:form_eq-1}
\Delta H_f({\rm X^{\rm q}}, \, \{\mu_i^\text{X}\}, \, E_\text{F}^*)  = \Delta H_f({\rm X^{\rm q^{\prime} }}, \, \{\mu_i^\text{X}\}, \, E_\text{F}^*).
\end{equation}
Using Eq. (\ref{eq:form_eq-1.0}), we observe that
\begin{equation}
\label{eq:form_eq-1.2}
\Delta H_f({\rm X^{\rm q}}, \, \{\mu_i^\text{X}\}, \, E_\text{F}) = \Delta H_f({\rm X^{\rm q}}, \, \{\mu_i^\text{X}\}, \, 0) + {\rm q}\, E_{\text{F}}.
\end{equation}
Given Eq. (\ref{eq:form_eq-1}) we can then employ Eq. (\ref{eq:form_eq-1.2}) to solve for $E_\text{F}^*$ which yields
\begin{equation}
\label{eq:CTL_eq-2} 
\epsilon^{\rm{X}} (\text{q}/\text{q}^{\prime}) \equiv E_\text{F}^* = \frac{\Delta H_f({\text{X}^{\rm q}}, \, \{\mu_i^\text{X}\}, \, 0)-\Delta H_f({\text{X}^{\rm q^{\prime} }}, \, \{\mu_i^\text{X}\}, \, 0)}
{\text{q}^{\prime} - \text{q}}.
\end{equation}
Upon using Eq. (\ref{eq:form_eq-1.0}) we can find an explicit expression for our ACTL~\cite{Freysoldt2014first},
\begin{equation}
\label{eq:CTL_eq-3} 
\epsilon^{\rm{X}} (\text{q}/\text{q}^{\prime}) =
\frac{(E_{\text{def}}({\rm X^{\rm q}}) +E_{\text{corr}}({\rm X^{\rm q}})) - (E_{\text{def}}({\rm X^{\rm q^{\prime}}}) +E_{\text{corr}}({\rm X^{\rm q^{\prime}}}))}{\text{q}^{\prime} - \text{q}}.
\end{equation} 
Here adiabatic simply reflects the fact that all formation energies that enter into the calculation of the charge-transition level are computed for defects that have been relaxed to their ground state.

% In considering the system of the two defect species,
% we assume that 

% \begin{align}
%     \left(H_f({\rm D^{\text{q}_\text{D}}}, \, \{\mu_i^\text{D}\}, \, E_{\text{F}'}) - \max_{\text{X}^\text{q} = \text{D}^0,\text{D}^+,\text{A}^-,\text{A}^0}H_f({\rm X^{\rm q}}, \, \{\mu_i^\text{X}\}, \, E_{\text{F}'})\right){\bigg/}k_BT \gg 1 \quad\forall\text{q}_\text{D} \neq 0,+1\label{eq:formen_cond1}\\
%     \left(H_f({\rm A^{\text{q}_\text{A}}}, \, \{\mu_i^\text{A}\}, \, E_{\text{F}'}) - \max_{\text{X}^\text{q} = \text{D}^0,\text{D}^+,\text{A}^-,\text{A}^0}H_f({\rm X^{\rm q}}, \, \{\mu_i^\text{X}\}, \, E_{\text{F}'})\right){\bigg/}k_BT \gg 1 \quad\forall\text{q}_\text{A} \neq 0,-1\label{eq:formen_cond2}
% \end{align} 

% Where $E_{\text{F}^{\prime} }$ is the value of 
% $E_{\text{F}}$ that satisfies charge conservation and equality of $n_{\text{D}}$ and $n_{\text{A}}$ at some fixed constant value for $n_{\text{D}}$ and $n_{\text{A}}$. If those conditions are satisfied, we can treat any donor-acceptor system as having a single ACTL, $\epsilon^{\rm{D}}(0/+)$, for the donor and a single ACTL, $\epsilon^{\rm{A}}(0/-)$, for the acceptor.

\subsection{Motivation for our Theoretical Formalism}
\label{sec:motivation}
Broadway \textit{et al.}~\cite{Broadway2018spat} determined the average electric field measured at the location of N$V^-$ defects within their sample by assuming the existence of a single equilibrium value for the Fermi level, $E_\text{F}(z)$, at every depth $z$ within the sample. We argue that the assumption of a single equilibrium value for $E_\text{F}$ throughout constant-$z$ planes in the theoretical formalism of Broadway \textit{et al.}~\cite{Broadway2018spat} may not hold. Specifically, we argue that over the duration of each measurement in the experiment of Broadway \textit{et al.} $E_\text{F}$ will only have time to equilibrate between no more than two defects, similar to an argument made by Collins~\cite{Collins2002}. We make this statement rigorous by determining the timescale for at least one acceptor to receive charge from at least one donor, which will set a lower bound on the timescale for $E_\text{F}$ to equilibrate in the sample. Ultimately, for $E_\text{F}$ to be at some constant equilibrium value over some region of the sample over the course of an experimental measurement, the timescale over which the charge in that region of the sample equilibrates must be much shorter than the timescale over which the experimental measurement is performed. 

Suppose we start from a time immediately after some perturbation to the sample has occurred such that an equilibrium value of $E_\text{F}$ must be re-established. If we are at the location of a donor at precisely that time and the donor is on the verge of transferring charge, $E_\text{F}$ will be pinned at the donor charge-transition level since the donor has not had time to interact with the rest of the sample. In general, a dopant can only contribute to charge equilibration in the sample if charge from the dopant has time to enter the conduction or valence band and travel to another defect. In the case of a donor, an electron from the donor must have time to reach the conduction band and travel to another defect. 

The calculation of the amount of time necessary for the transfer of charge proceeds as follows. The expectation value for the speed at which an electron can travel in the crystal once in the conduction band is given by~\cite{Kaxiras2003atomic}
\begin{equation}
    \left<s_\mathbf{k}\right> = \frac{1}{\hbar}|\nabla_\mathbf{k}\epsilon^\text{C}_{\mathbf{k},s}|.
\end{equation}
Above, $|\nabla_\mathbf{k}\epsilon^\text{C}_{\mathbf{k},s}|$ denotes the norm of the gradient with respect to the wave-vector $\mathbf{k}$ of some conduction-band eigenvalue with spin $s$ evaluated at $\mathbf{k}$ and $\hbar$ is the reduced Planck constant. 

If $E_\text{F}$ lies below $\epsilon^\text{C}_{\mathbf{k},s}$ for all $\mathbf{k}$, the band will only be occupied by a donor electron a fraction of the time given by the Fermi-Dirac distribution with local Fermi level $E_\text{F}$, which will be reflected in the expectation value for the speed. After the electron has traveled for a time $t-t_0$ starting at a time $t_0$, the acceptor will only be able to receive an electron from the conduction band state at wavevector $\mathbf{k}^\prime$ given by $\mathbf{k}^\prime = \mathbf{k}(t_0) +\frac{1}{\hbar}\int_{t_0}^t\mathbf{F}_e(\tilde{t}^\prime)\text{d}\tilde{t}^\prime$ a fraction of the time given by 1 minus the Fermi-Dirac distribution with the acceptor's local Fermi level $E_{\text{F}}^\prime$. Above, $\mathbf{F}_e(\tilde{t}^\prime)$ is the external force acting on the electron. As a result of the external force $\mathbf{F}_e(\tilde{t}^\prime)$, we must introduce a time dependence to $\mathbf{k}$. Explicitly, averaging over time, $\left<s_\mathbf{k}\right> = \left<s_{\mathbf{k},\mathbf{k}^\prime}\right>$ becomes
\begin{align}
    \left<s_{\mathbf{k},\mathbf{k}^\prime}\right> &= \frac{1}{t-t_0}\int_{t_0}^t\text{d}t^\prime\frac{1}{\hbar}|\nabla_\mathbf{k}\epsilon^\text{C}_{\mathbf{k}(t^\prime),s}|\times\frac{1}{\exp((\epsilon^\text{C}_{\mathbf{k}(t_0),s}-E_\text{F})/k_BT)+1}\\&\times\frac{\delta_{\mathbf{k}^\prime,\mathbf{k}(t_0)+\frac{1}{\hbar}\int_{t_0}^{t}\mathbf{F}_e(\tilde{t}^\prime)\text{d}\tilde{t}^\prime}}{\exp((E_{\text{F}}^\prime-\epsilon^\text{C}_{\mathbf{k}^\prime,s})/k_BT)+1}.\nonumber
\end{align}
 The expected rate at which a donor can transfer an electron from its location to an acceptor at another location is then given by
\begin{align}
    \left<\Gamma_{\mathbf{k},\mathbf{k}^\prime}\right> &= \int_{t_0}^\infty\text{d}t\frac{|\partial_t(|\Delta\mathbf{r}|^2-\int_{t_0}^t\text{d}\tilde{t}\frac{1}{\hbar}\nabla_\mathbf{k}\epsilon^\text{C}_{\mathbf{k}(\tilde{t}),s}\cdot\Delta\mathbf{r})|}{t-t_0}\int_{t_0}^t\text{d}t^\prime\frac{1}{\hbar}\frac{\nabla_\mathbf{k}\epsilon^\text{C}_{\mathbf{k}(t^\prime),s}\cdot\Delta\mathbf{r}}{|\Delta\mathbf{r}|^2}\\&\times\frac{\delta(|\Delta\mathbf{r}|^2-\int_{t_0}^t\text{d}\tilde{t}\frac{1}{\hbar}\nabla_\mathbf{k}\epsilon^\text{C}_{\mathbf{k}(\tilde{t}),s}\cdot\Delta\mathbf{r})}{\exp((\epsilon^\text{C}_{\mathbf{k}(t_0),s}-E_\text{F})/k_BT)+1}\times\frac{\delta_{\mathbf{k}^\prime,\mathbf{k}(t_0)+\frac{1}{\hbar}\int_{t_0}^t\mathbf{F}_e(\tilde{t}^\prime)\text{d}\tilde{t}^\prime}}{\exp((E_{\text{F}}^\prime-\epsilon^\text{C}_{\mathbf{k}^\prime,s})/k_BT)+1},\nonumber
\end{align}
where $E_\text{F}$ is pinned at the donor level, $E_\text{F}^\prime$ is pinned at the acceptor level, and $\Delta\mathbf{r}$ is the displacement from the donor to the acceptor.

Assuming interactions between acceptors and donors in the sample occur randomly and independently, summing such expressions over all donors and all acceptors in the sample and over $\mathbf{k}^\prime$ and averaging over all displacements between defects as well as over $\mathbf{k}(t_0)$ and $s$ gives the effective rate for the transfer of electrons which takes the form
\begin{align}
\label{eq:eGamma}
    \bar{\Gamma}_e &= \frac{1}{\hbar}\sum_{\text{D}~\in~\text{Donors}}N_{\text{D}}\int_{V_\text{D}}\text{d}\mathbf{r}\rho_\text{D}(\mathbf{r})\sum_{\text{A}~\in~\text{Acceptors}}N_{\text{A}}\int_{V_\text{A}}\text{d}\mathbf{r}^\prime\rho_\text{A}(\mathbf{r}^\prime)\frac{1}{N_s}\sum_{s}\Omega_{PUC}\int\frac{\text{d}\mathbf{k}(t_0)}{(2\pi)^3}\int\frac{\text{d}\mathbf{k}^\prime}{(2\pi)^3}\\&\int_{t_0}^\infty\text{d}t\frac{|\partial_t(|\mathbf{r}^\prime-\mathbf{r}|^2-\int_{t_0}^t\text{d}\tilde{t}\frac{1}{\hbar}\nabla_\mathbf{k}\epsilon^\text{C}_{\mathbf{k}(\tilde{t}),s}\cdot(\mathbf{r}^\prime-\mathbf{r}))|}{t-t_0}\int_{t_0}^t\text{d}t^\prime\frac{\nabla_\mathbf{k}\epsilon^\text{C}_{\mathbf{k}(t^\prime),s}\cdot(\mathbf{r}^\prime-\mathbf{r})}{|\mathbf{r}^\prime-\mathbf{r}|^2}\nonumber\\&\times\frac{\delta(|\mathbf{r}^\prime-\mathbf{r}|^2-\int_{t_0}^t\text{d}\tilde{t}\frac{1}{\hbar}\nabla_\mathbf{k}\epsilon^\text{C}_{\mathbf{k}(\tilde{t}),s}\cdot(\mathbf{r}^\prime-\mathbf{r}))}{\exp((\epsilon^\text{C}_{\mathbf{k}(t_0),s}-E_{\text{F}})/k_BT)+1}\nonumber\times\frac{\delta(\mathbf{k}^\prime-\mathbf{k}(t_0)-\frac{1}{\hbar}\int_{t_0}^t\mathbf{F}_e(\tilde{t}^\prime)\text{d}\tilde{t}^\prime)}{\exp((E_{\text{F}}^\prime-\epsilon^\text{C}_{\mathbf{k}^\prime,s})/k_BT)+1}\nonumber
\end{align}
In Eq. (\ref{eq:eGamma}), $N_\text{D}$ and $N_\text{A}$ are the total numbers of D and A defects in the entire crystal. In order to avoid artificially reducing the defect concentrations, the integration with respect to $\mathbf{r}$ is performed over a region containing one donor on average ($V_\text{D} = 1/n_\text{D}$) and the integration with respect to $\mathbf{r}^\prime$ is performed over a region containing one acceptor on average ($V_\text{A} = 1/n_\text{A}$). We employ cubes with centers at zero for the integrations. The quantities $\rho_\text{D}(\mathbf{r})$ and $\rho_\text{A}(\mathbf{r}^\prime)$ are the concentrations of species D and A as functions of position. The integer $N_s$ represents the number of spin states and $\Omega_{PUC}$ represents the volume of the primitive unit cell of the fcc lattice of diamond. Averaging over displacements as well as over $\mathbf{k}(t_0)$ and $s$ is performed due to the fact that a single defect cannot be measured as simultaneously having multiple positions and a single charge cannot be measured as simultaneously having multiple speeds or spins.

Equivalently, for the transfer of holes we have the effective rate
\begin{align}
\label{eq:hGamma}
    \bar{\Gamma}_h &= \frac{1}{\hbar}\sum_{\text{D}~\in~\text{Donors}}N_{\text{D}}\int_{V_\text{D}}\text{d}\mathbf{r}\rho_\text{D}(\mathbf{r})\sum_{\text{A}~\in~\text{Acceptors}}N_{\text{A}}\int_{V_\text{A}}\text{d}\mathbf{r}^\prime\rho_\text{A}(\mathbf{r}^\prime)\frac{1}{N_s}\sum_{s}\Omega_{PUC}\int\frac{\text{d}\mathbf{k}(t_0)}{(2\pi)^3}\int\frac{\text{d}\mathbf{k}^\prime}{(2\pi)^3}\\&\int_{t_0}^\infty\text{d}t\frac{|\partial_t(|\mathbf{r}-\mathbf{r}^\prime|^2-\int_{t_0}^t\text{d}\tilde{t}\frac{1}{\hbar}\nabla_\mathbf{k}\epsilon^\text{V}_{\mathbf{k}(\tilde{t}),s}\cdot(\mathbf{r}-\mathbf{r}^\prime))|}{t-t_0}\int_{t_0}^t\text{d}t^\prime\frac{\nabla_\mathbf{k}\epsilon^\text{V}_{\mathbf{k}(t^\prime),s}\cdot(\mathbf{r}-\mathbf{r}^\prime)}{|\mathbf{r}-\mathbf{r}^\prime|^2}\nonumber\\&\times\frac{\delta(|\mathbf{r}-\mathbf{r}^\prime|^2-\int_{t_0}^t\text{d}\tilde{t}\frac{1}{\hbar}\nabla_\mathbf{k}\epsilon^\text{V}_{\mathbf{k}(\tilde{t}),s}\cdot(\mathbf{r}-\mathbf{r}^\prime))}{\exp((\epsilon^\text{V}_{\mathbf{k}^\prime,s}-E_{\text{F}})/k_BT)+1}\nonumber\times\frac{\delta(\mathbf{k}^\prime-\mathbf{k}(t_0)-\frac{1}{\hbar}\int_{t_0}^t\mathbf{F}_h(\tilde{t}^\prime)\text{d}\tilde{t}^\prime)}{\exp((E_{\text{F}}^\prime-\epsilon^\text{V}_{\mathbf{k}(t_0),s})/k_BT)+1}\nonumber.
\end{align}
where $\epsilon^\text{V}_{\mathbf{k},s}$ denotes some valence-band eigenvalue with wavevector $\mathbf{k}$ and spin $s$ and $\mathbf{F}_h$ is the external force acting on the holes. The desired total effective rate is simply 
\begin{equation}
\label{eq:totalGamma}
    \bar{\Gamma} = \bar{\Gamma}_e+\bar{\Gamma}_h.
\end{equation}
The reciprocal of the effective rate gives the desired timescale for the equilibration of $E_\text{F}$. Below, we will consider that the defects being measured are sufficiently deep in the bulk that the forces from the randomly distributed charged defects surrounding them average to zero. We also note that for an equilibrium reaction the rates of the forward and reverse reactions must be equal. Therefore, whether we calculate the charge transfer rate for the case where a defect pair is initially neutral and becomes ionized or for the case where a defect pair is initially ionized and becomes neutral does not affect the final result. For ease of calculation, we consider the case where the defect pair being measured is initially neutral and becomes ionized. 

We now turn to the computational evaluation of Eq. (\ref{eq:totalGamma}). We make the approximation that the external forces on the electrons and on the holes are random and integrate to zero between any two times $t_0$ and $t$ and we use the fact that for the N$V$ acceptor and N$_\text{C}$ donor $(E_{\text{F}}^\prime-\epsilon^\text{V}_{\mathbf{k}^\prime,s}) - (\epsilon^\text{C}_{\mathbf{k},s}-E_\text{F}) \gg k_BT, E_\text{F}-E_{\text{F}}^\prime \gg k_BT~\forall \mathbf{k},\mathbf{k}^\prime$. We further assume an isotropic distribution of defects. In our isotropic model, at every distance $r$ from some origin, the concentration of X defects between that distance and a distance infinitesimally farther will be $\frac{\tilde{N}_\text{X}}{4\pi r^2\text{d}r}$, given $\tilde{N}_\text{X}$ defects in the spherical shell of interest. Since $\tilde{N}_\text{X} = 4\pi r^2\text{d}r\cdot n_\text{X}$, the expression for the concentration in a shell ensures that isotropy is satisfied. As alluded to above, we must integrate to a distance such that one defect is enclosed in the entire region. For the species X, $\text{d}r \approx (3V_{\text{X}}/(4\pi))^{1/3}$. We then have
\begin{align}
\label{eq:eGammaapprox}
    \bar{\Gamma}_e &\approx \frac{1}{\hbar}\sum_{\text{D}~\in~\text{Donors}}\frac{N_{\text{D}}}{(3V_{\text{D}}/(4\pi))^{1/3}}\int_{V_\text{D}}\text{d}\mathbf{r}\frac{1}{4\pi r^2}\sum_{\text{A}~\in~\text{Acceptors}}\frac{N_{\text{A}}}{(3V_{\text{A}}/(4\pi))^{1/3}}\int_{V_\text{A}}\text{d}\mathbf{r}^\prime\frac{1}{4\pi {r^\prime}^2}\frac{1}{N_s}\sum_{s}\Omega_{PUC}\times\\&\int\frac{\text{d}\mathbf{k}}{(2\pi)^3}\frac{|\nabla_\mathbf{k}\epsilon^\text{C}_{\mathbf{k},s}|}{|\mathbf{r}^\prime-\mathbf{r}|}\times\frac{1}{\exp((\epsilon^\text{C}_{\mathbf{k},s}-E_\text{F})/k_BT)+1}\nonumber
\end{align}
and
\begin{equation}
\label{eq:hGammaapprox}
    \bar{\Gamma}_h \approx 0.
\end{equation}
The increment in $\mathbf{k}$ in the integral over $\mathbf{k}$ is constrained by the size of the crystal. Given our desire to capture only leading-order contributions, we considered only the N$_\text{C}$ donor and the N$V$ acceptor in our rate calculations.  Due to the exponential suppression of the corresponding contribution to the rate if $\epsilon^\text{C}_{\mathbf{k},s} - E_\text{F} \gg k_BT$, for our computations we iteratively considered $\epsilon^\text{C}_{\mathbf{k},s} \leq E_\text{C}+mk_BT$ for increasing integers $m$, where $E_\text{C}$ is the energy of the conduction band minimum. We found that the change in the rate was less than $3$\% between $m=7$ and $m=8$ for all k-point resolutions at $T = 300$~K and less than $1$\% between $m=7$ and $m=8$ for the three highest k-point resolutions at that temperature. We also observed convergence of the rate to within $3$\% between the two highest k-point resolutions. We therefore did not investigate $m > 8$ or a k-point resolution higher than 200 k-points between consecutive high-symmetry points. In order to integrate over the volume surrounding the location of the conduction band minimum in the Brillouin zone, we have assumed isotropic dispersion near the conduction band minimum. 

\begin{figure}[ht!] 
\centering
\includegraphics[width=0.91\textwidth]{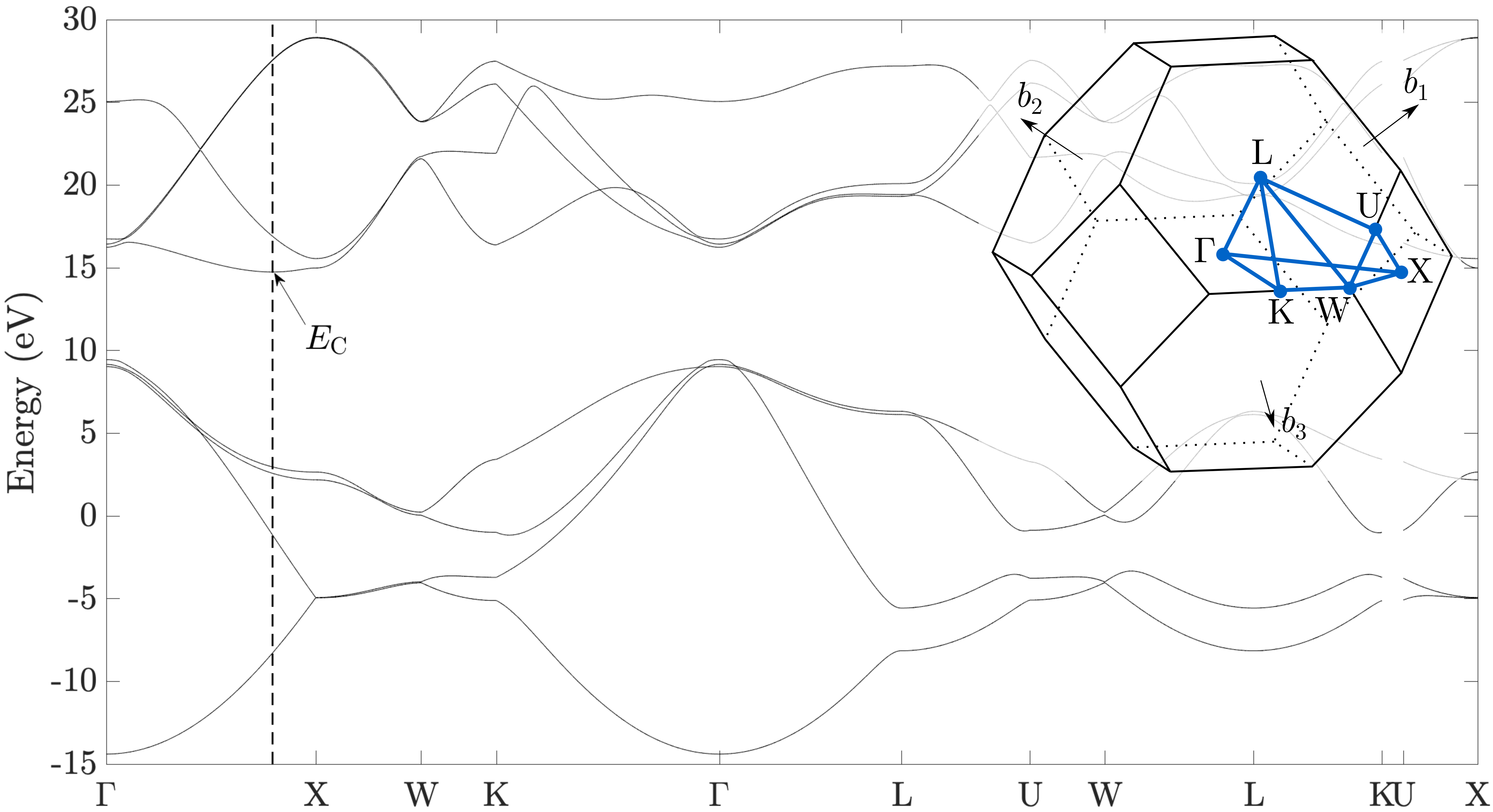}
\caption{Spin-polarized bandstructure for the primitive fcc unit cell of diamond at a resolution of 200 k-points along the lines connecting consecutive high-symmetry points in the bandstructure path. The position of the conduction band minimum with energy $E_\text{C}$ is indicated. The calculated band gap is $E_\text{C}-E_\text{V} \approx 5.3$~eV, where $E_\text{V}$ is the energy of the valence band maximum. The first fcc Brillouin zone is depicted in the upper-right quarter of the figure along with its high-symmetry points and the primitive vectors of the reciprocal lattice $b_1 = \frac{2\pi}{a}(-\hat{k}_x+\hat{k}_y+\hat{k}_z)$, $b_2 = \frac{2\pi}{a}(\hat{k}_x-\hat{k}_y+\hat{k}_z)$, and $b_3 = \frac{2\pi}{a}(\hat{k}_x+\hat{k}_y-\hat{k}_z)$.} 
\label{fig:bandstructure}
\end{figure}

\begin{figure}[ht!] 
\centering
\includegraphics[width=0.6\textwidth]{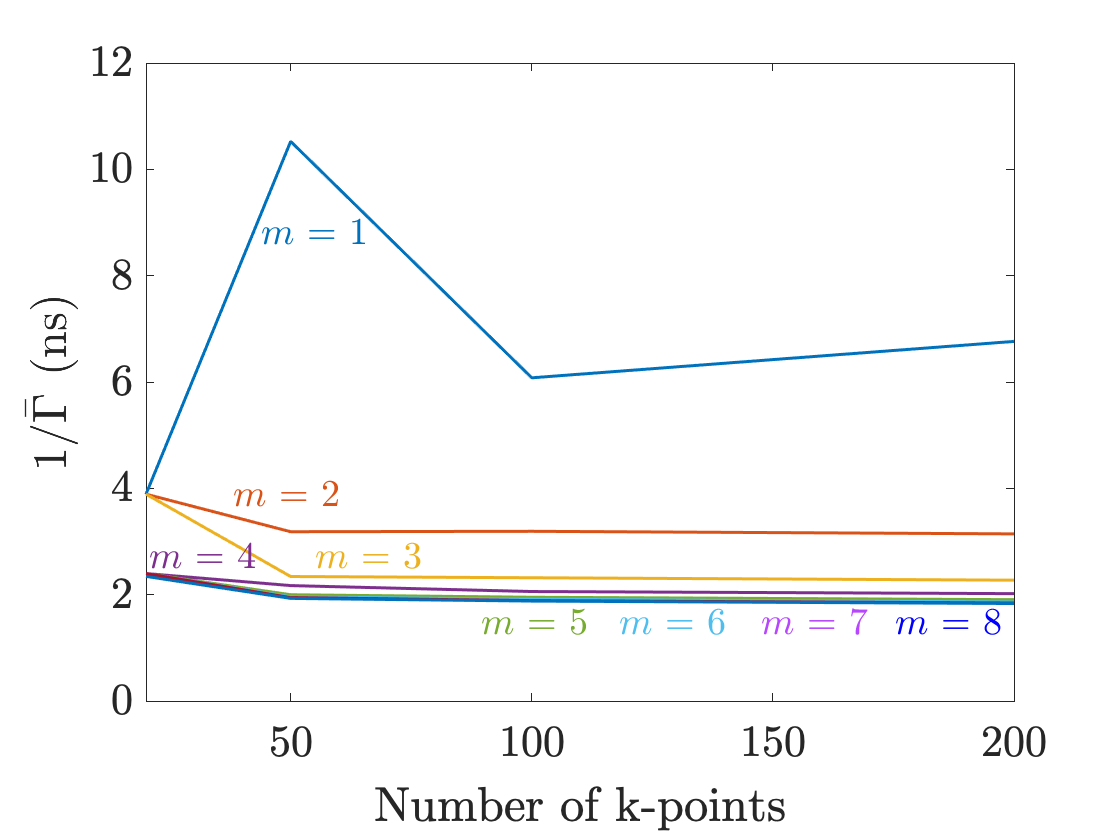}
\caption{Minimum timescale in nanoseconds for the equilibration of $E_\text{F}$ for various cutoffs $\epsilon^\text{C}_\mathbf{k} \leq E_\text{C} +mk_BT$ as a function of the number of k-points along the lines connecting consecutive high-symmetry points in the bandstructure path (see Fig. \ref{fig:bandstructure}). Timescales were computed for 20, 50, 100, and 200 k-points along the lines connecting consecutive high-symmetry points in the bandstructure path.} 
\label{fig:rateinv}
\end{figure}

As alluded to above, we considered only the N$_\text{C}$ donor and the N$V$ acceptor in our rate calculations. Therefore, $V_\text{D} = V_{\text{N}_\text{C}} = 1/n_{\text{N}_\text{C}}$ and $V_\text{A} = V_{\text{N}V} = 1/n_{\text{N}V}$. The converged lattice constant for the primitive fcc unit cell of diamond was $a=3.549$ \AA. This value is in good agreement with a previous theoretical calculation of $a=3.545$~\AA~\cite{Deak2014formation}. Thus, the number of reciprocal lattice points along a line connecting two high-symmetry points that is physically needed to resolve the increment in $\mathbf{k}$ is approximately 200. At the highest k-point resolution, we found an electronic band gap of 5.3~eV, also in good agreement with previous experimental~\cite{Madelung1991semiconductors} and theoretical~\cite{Deak2014formation} results. Given the high nitrogen fluence used in the Broadway \textit{et al.} experiment, we employed $\chi = 0.01$ for the N$V$ yield with $n_\text{D} = n_{\text{N}_\text{C}} \approx 1.41\times10^{18}$~cm$^{-3}$ so that $n_\text{A} = n_{\text{N}V} \approx 1.43\times10^{16}$~cm$^{-3}$ using $\chi n_{\text{N}_\text{C}} = (1-\chi)n_{\text{N}V}$~\cite{Broadway2018spat}. The bandstructure for the highest k-point resolution with the position of the conduction band minimum indicated is depicted in Fig. \ref{fig:bandstructure}. For various $m$, the quantity $1/\bar{\Gamma}$ as a function of the number of k-points along the lines connecting consecutive high-symmetry points in the bandstructure path is provided in Fig. \ref{fig:rateinv}. As shown in Fig. \ref{fig:rateinv}, the rate converges to a value corresponding to a minimum timescale for the equilibration of $E_\text{F}$ that is approximately equal to 1.8~ns. By comparison, the timescale over which the measurements in the experiment of Broadway \textit{et al.} were performed was approximately 11 microseconds~\cite{Broadway2018spat}. For $n_{\text{N}V} \approx 1.43\times10^{16}$~cm$^{-3}$, there exist approximately 4 billion N$V$ defects in the 2~mm $\times$ 2~mm $\times$ 70~nm portion of the sample. Therefore, the probability that a defect would not have reached equilibrium with more than a single other defect after 11 microseconds is greater than $(1-\frac{1}{4\times10^9})^{11~\mu\text{s}/1.8~\text{ns}} \approx 0.9999985$. As a consequence, the vast majority of N$V$ defects in the sample will not have reached equilibrium with more than a single other defect after 11 microseconds. Considering contributions to the rate from the entire sample of size 2 mm $\times$ 2 mm $\times$ 50 $\mu$m leads to no appreciable change in the rate using the fact that $n_{\text{N}_\text{C}} < 1$~ppb in the rest of the sample~\cite{Broadway2018spat} and keeping $\chi = 0.01$. If 1.8~ns is the shortest duration for equilibration between any pair of N$_\text{C}$ and N$V$ defects in the entire sample, we must conclude that in the vast majority of cases $E_\text{F}$ will at most be in equilibrium between pairs of N$_\text{C}$ and N$V$ defects in each 11 microsecond measurement in the experiment of Broadway \textit{et al.}

Given the relatively long timescales required for the equilibrium transfer of charge between the ground-state defect levels, photoexcitation or photoionization would be necessary in order to perform experiments with charged defects on reasonable timescales~\cite{Broadway2018spat}. Nonetheless, as the time required for the charge to equilibrate thermally would still be dictated by our formalism and since thermal equilibrium is required for the applicability of the concept of a uniform Fermi level, our analysis should not be affected by the use of optical illumination. Our results could also help explain the need for the trap-filling procedure by photoexcitation used in highly pure radiation detectors~\cite{Lee1999compensation}. Additionally, our formalism could help quantify charge-transfer rates resulting from the placement of electrons in excited states as a result of laser-induced perturbation for charge transport between individual fluorescent defects~\cite{Lozovoi2021optical}.

\subsection{Overview of our Theoretical Formalism}
Given the conclusion that $E_\text{F}$ could at most be in equilibrium between N$_\text{C}$ and N$V$ defect pairs in the experiment of Broadway \textit{et al.}, consider the case where in some subregion of the crystal sample the total defect concentrations $n_\text{D} = n_\text{A} = C$ for some constant $C$ such that the subregion can be considered to consist of a single donor equilibrating with a single acceptor. We assume that the donor and acceptor are initially neutral before they exchange charge to equilibrate with one another so that $n_{\text{D}^{0}} = n_{\text{A}^{0}}$, which is achieved mathematically by appropriately setting $\{\mu_i^\text{D}\}$ and $\{\mu_i^\text{A}\}$ in Eq. (\ref{eq:concentration}). We make the additional assumption that the concentrations of the charge states other than $\text{D}^{0}$, $\text{D}^{+}$, $\text{A}^{0}$, and $\text{A}^{-}$ are suppressed in thermal equilibrium. This additional assumption implies that $(\epsilon^{\rm{D}}(0/\text{q}_\text{D}) - \epsilon^{\rm{A}}(0/-)) \gg k_BT$, $(\epsilon^{\rm{D}}(0/+) - \epsilon^{\rm{A}}(0/\text{q}_\text{A})) \gg k_BT$, $(\epsilon^{\rm{A}}(0/\text{q}'_\text{A}) - \epsilon^{\rm{A}}(0/-)) \gg k_BT$, and $(\epsilon^{\rm{D}}(0/+) - \epsilon^{\rm{D}}(0/\text{q}'_\text{D})) \gg k_BT   ~\forall\text{q}_\text{D},\text{q}_\text{A},\text{q}'_\text{A},\text{q}'_\text{D},~\text{s.t.}~ \text{q}_\text{D} < 0,~ \text{q}_\text{A} > 0,~\text{q}'_\text{A} < -1$, and $~ \text{q}'_\text{D} > +1$. Physically, the conditions mean that the value of $E_\text{F}$ required to transition to any charge state of D or A other than $\text{D}^{0}$, $\text{D}^{+}$, $\text{A}^{0}$, and $\text{A}^{-}$ must be much higher (lower) than the value of $E_\text{F}$ that maximizes the minimum of $n_{\text{D}^{+}}$ and $n_{\text{A}^{-}}$ when the other charge state is negative (positive) and when $n_{\text{D}^{0}} = n_{\text{A}^{0}}$, which implies that the other charge state will be suppressed if $E_\text{F}$ is indeed chosen to maximize the minimum of $n_{\text{D}^{+}}$ and $n_{\text{A}^{-}}$ when $n_{\text{D}^{0}} = n_{\text{A}^{0}}$. We further assume that the defect levels are separated from the band edges by an amount much greater than $k_BT$ so that the electron and hole concentrations ($n$ and $p$, respectively) satisfy $n \approx 0$ and $p \approx 0$.   
% If the conditions are met, a charge-conserving solution can be obtained by choosing $E_{\rm F}$, $\{\mu^\text{D}_i\}$, and $\{\mu^\text{A}_i\}$ such that $n_{\text{D}^+} = n_{\text{A}^-}$ and $n_\text{D} = n_\text{A} = C$. The conditions in Eqs. (\ref{eq:formen_cond1}) and (\ref{eq:formen_cond2})
% will ensure that the concentrations of the charge states other than D$^-$, D$^0$, A$^+$, and A$^0$ are suppressed at the value of $E_\text{F}$ that satisfies charge conservation and $n_\text{D} = n_\text{A} = C$. The additional condition that the levels of both defects be separated from the band edges by an amount much greater than $k_BT$ implies $n_0 \approx p_0 \approx 0$, where $n_0$ and $p_0$ are the electron and hole carrier concentrations, respectively. 
Therefore, allowing the total concentration of donors to be equal to the total concentration of acceptors and imposing charge conservation we can then write to a good approximation
\begin{align}
\label{eq:eqconc}
   n_{\text{D}^{+}}+n_{\text{D}^{0}} = n_{\text{A}^{-}}+n_{\text{A}^{0}} = C,\\
\label{eq:chgbalance}
   n_{\text{D}^{+}} = n_{\text{A}^{-}}.
\end{align}

We now proceed to determine the equilibrium value of $E_\text{F}$ in this system of two defects. Starting from Eq. (\ref{eq:chgbalance}), we can use Eq. (\ref{eq:concentration}) to obtain
\begin{equation}
\label{eq:deriv1}
    N_{\text{D}}g_{\text{D}^+}\exp(-\Delta H_f(\text{D}^+,\{\mu_i^\text{D}\},E_\text{F})/k_BT) = N_{\text{A}}g_{\text{A}^-}\exp(-\Delta H_f(\text{A}^-,\{\mu_i^\text{A}\},E_\text{F})/k_BT).
\end{equation}
Therefore, using Eq. (\ref{eq:form_eq-1.2}), Eq. (\ref{eq:deriv1}) becomes
\begin{align}
    &N_{\text{D}}g_{\text{D}^+}\exp(-\Delta H_f(\text{D}^+,\{\mu_i^\text{D}\},0)/k_BT)\times\exp(-E_\text{F}/k_BT)\\ &= N_{\text{A}}g_{\text{A}^-}\exp(-\Delta H_f(\text{A}^-,\{\mu_i^\text{A}\},0)/k_BT)\times\exp(E_\text{F}/k_BT).\nonumber
\end{align}
Solving for $E_\text{F}$ we have
\begin{equation}
\label{eq:chgconsFermi}
    E_\text{F} = (\Delta H_f({\rm A^{-}}, \, \{\mu_i^\text{A}\}, \, 0) - \Delta H_f({\rm D^{+}}, \, \{\mu_i^\text{D}\}, \, 0))/2 + \frac{k_BT}{2}\ln\left(\frac{N_\text{D}g_{\text{D}^+}}{N_\text{A}g_{\text{A}^-}}\right).
\end{equation}
Next, we use Eq. (\ref{eq:chgbalance}) to eliminate $n_{\text{D}^+}$ and $n_{\text{A}^-}$ from Eq. (\ref{eq:eqconc}) which recovers
\begin{equation}
\label{eq:deriv2}
  n_{\text{D}^{0}} = n_{\text{A}^{0}}.
\end{equation}
We can again use Eq. (\ref{eq:concentration}) to obtain
\begin{equation}
\label{eq:deriv3}
    N_{\text{D}} g_{\text{D}^{0}} \exp(-\Delta H_f(\text{D}^{0},\{\mu_i^\text{D}\},E_\text{F})/k_BT) = N_{\text{A}}g_{\text{A}^{0}}\exp(-\Delta H_f(\text{A}^0,\{\mu_i^\text{A}\},E_\text{F})/k_BT).
\end{equation}
Using Eq. (\ref{eq:form_eq-1.2}), this expression becomes
\begin{equation}
\label{eq:deriv4}
    N_{\text{D}}g_{\text{D}^0}\exp(-\Delta H_f(\text{D}^0,\{\mu_i^\text{D}\},0)/k_BT) = N_{\text{A}}g_{\text{A}^0}\exp(-\Delta H_f(\text{A}^0,\{\mu_i^\text{A}\},0)/k_BT),
\end{equation} 
which can be rearranged to give
\begin{equation}
\label{eq:deriv5}
\Delta H_f(\text{A}^{0}, \{\mu_i^\text{A}\}, \, 0) -\Delta H_f(\text{D}^{0}, \{\mu_i^\text{D}\}, \, 0) +k_BT\ln\left(\frac{N_{\text{D}}g_{\text{D}^0}}{N_{\text{A}}g_{\text{A}^0}}\right) = 0.
\end{equation}
If we divide Eq. (\ref{eq:deriv5}) by a factor of 2 and combine the result with Eq. (\ref{eq:chgconsFermi}) we obtain

\begin{align}
\label{eq:key-equation}
    E_\text{F} &= (\Delta H_f({\rm A^{-}}, \, \{\mu_i^\text{A}\}, \, 0) -\Delta H_f({\rm A^{0}}, \, \{\mu_i^\text{A}\}, \, 0))/2\\ &+(\Delta H_f({\rm D^{0}}, \, \{\mu_i^\text{A}\}, \, 0) -\Delta H_f({\rm D^{+}}, \, \{\mu_i^\text{D}\}, \, 0))/2\nonumber\\ &+\frac{k_BT}{2}\ln\left(\frac{g_{\text{D}^+}g_{\text{A}^0}}{g_{\text{D}^0}g_{\text{A}^-}}\right)\nonumber
    \end{align}
Using Eq. (\ref{eq:CTL_eq-2}) we obtain
\begin{equation}
\label{eq:equilFermi0-1.1}
E_\text{F} = \frac{\epsilon^{\rm{D}} (0/+)+\epsilon^{\rm{A}} (0/-)}{2} + \frac{k_BT}{2}\ln\left(\frac{g_{\text{D}^{+}}g_{\text{A}^{0}}}{g_{\text{D}^{0}}g_{\text{A}^{-}}}\right).
\end{equation}
Dropping the contribution from the configurational entropy, we therefore define an ACTL for one defect in the presence of another given by
\begin{equation}
\label{eq:chgtransitionlevel}
    \epsilon^{\rm{D},\rm{A}} (0/+,0/-) \equiv \frac{\epsilon^{\rm{D}} (0/+)+\epsilon^{\rm{A}} (0/-)}{2}.
\end{equation}

A benefit of the form of Eq. (\ref{eq:chgtransitionlevel}) is that it is manifestly independent of the choice of $\{\mu_i^\text{D}\}$ and $\{\mu_i^\text{A}\}$. Also, we readily see that if $\{\mu_i^\text{D}\}$ and $\{\mu_i^\text{A}\}$ are chosen such that $n_{\text{D}^{0}} = n_{\text{A}^{0}}$, $E_\text{F} = \epsilon^{\rm{D},\rm{A}} (0/+,0/-)$ with the addition of a term of order $k_BT$ will indeed maximize the minimum of $n_{\text{D}^{+}}$ and $n_{\text{A}^{-}}$. To give some intuition for the expression in Eq. (\ref{eq:equilFermi0-1.1}), we note that Broadway \textit{et al.}~\cite{Broadway2018spat} investigated the situation after the donor (N$_\text{C}$) has lost its electron to the acceptor (N$V$) so that the donor and acceptor roles are reversed. Therefore, the value $E_\text{F} \approx \left(\epsilon^{\rm{N}_\text{C}}(0/+)+\epsilon^{{\rm N}V}(0/-)\right)/2$ parallels the result from solid-state theory that $E_\text{F}$ lies halfway between the electron accepting conduction band and the electron donating valence band for equal band curvatures.

From the energetics of the charge transfer, we determine the total amount of band bending. Using a generalization of the complex binding energy~\cite{Freysoldt2014first}, we had previously shown that the error associated with assuming the dilute limit when taking the average of the ACTLs is negligible compared to the averaged transition energy~\cite{Kuate2021theor}. Therefore, to a good approximation the energy required to ionize the system of two defects can be taken to be equal to the average of the dilute-limit ACTLs for the two defects. Such a result is consistent with the electronic structure of the semiconductor being modulated by the presence of the two defects in such a manner as to result in a band-bending profile along the line connecting them. We demonstrate this consistency by first noting that $E_\text{F}$ must be pinned at the acceptor and donor levels in the respective parts of the sample if A gains a single electron and D loses a single electron when they are sufficiently far apart that the dilute limit can be applied. In equilibrium, however, $E_\text{F}$ is constant throughout the subregion of the sample containing the two defects. Thus, the conduction band minimum ($E_\text{C}$) and the valence band maximum ($E_\text{V}$) must be shifted at the positions $\mathbf{r}_\text{A}$ and $\mathbf{r}_\text{D}$ of the respective defects,
\begin{equation}
\label{eq:CBMVBMshiftA}
    \Delta{E_\text{C}}(\mathbf{r}_\text{A}) =  \frac{1}{2}\left(\epsilon^{\rm{D}}(0/+)-\epsilon^{\rm{A}}(0/-)\right)  = \Delta{E_\text{V}}(\mathbf{r}_\text{A})
\end{equation}
and
\begin{equation}
\label{eq:CBMVBMshiftD}
    \Delta{E_\text{C}}(\mathbf{r}_\text{D}) =  \frac{1}{2}\left(\epsilon^{\rm{A}}(0/-)-\epsilon^{\rm{D}}(0/+)\right) = \Delta{E_\text{V}}(\mathbf{r}_\text{D}).
\end{equation}
 In traveling from the location of the donor defect to the location of the acceptor defect, we obtain the result for the total bending of the conduction and valence band extrema,
\begin{align}
    {E_\text{C}}(\mathbf{r}_\text{A})-{E_\text{C}}(\mathbf{r}_\text{D}) &= \Delta{E_\text{C}}(\mathbf{r}_\text{A})-\Delta{E_\text{C}}(\mathbf{r}_\text{D}) = \left(\epsilon^{\rm{D}}(0/+)-\epsilon^{\rm{A}}(0/-)\right),\\
    {E_\text{V}}(\mathbf{r}_\text{A})-{E_\text{V}}(\mathbf{r}_\text{D}) &= \Delta{E_\text{V}}(\mathbf{r}_\text{A})-\Delta{E_\text{V}}(\mathbf{r}_\text{D}) = \left(\epsilon^{\rm{D}}(0/+)-\epsilon^{\rm{A}}(0/-)\right).
\end{align} 

The electric field associated with the bending of the conduction and valence bands due to the presence of the defects is then given by~\cite{Zhang2012band,Broadway2018spat,Dalven1990introduction}
\begin{align}
\label{eq:Efield}
    \vec{\mathcal{E}} = -\frac{1}{(-e)}\nabla({E_\text{V}}(\mathbf{r})) \approx \frac{1}{e}\frac{({E_\text{V}}(\mathbf{r}_\text{A})-{E_\text{V}}(\mathbf{r}_\text{D}))}{|\Delta\mathbf{r}|}\frac{\Delta\mathbf{r}}{|\Delta\mathbf{r}|},
\end{align}
where, as above, $E_\text{V}$ is the valence band maximum, $e$ is the elementary charge, $\mathbf{r}_\text{A}$ is the position of defect A, $\mathbf{r}_\text{D}$ is the position of defect D, and $\Delta\mathbf{r} = \mathbf{r}_\text{A} - \mathbf{r}_\text{D}$.  
Considering the case of N$V$ defects in the presence of N$_\text{C}$ defects in diamond, we note that a diamond crystal containing N$V$ and N$_\text{C}$ defects satisfies the conditions required for the applicability of Eqs. (\ref{eq:eqconc}) and (\ref{eq:chgbalance})~\cite{Deak2014formation,Kuate2021theor}. Thus, Eq. (\ref{eq:Efield}) applies to N$V$ defects in the presence of N$_\text{C}$ defects in diamond.

 \section{DISCUSSION AND ELUCIDATION OF THE EXPERIMENT OF BROADWAY \textit{ET AL.} \label{sec:imp}}
\subsection{Details of the Broadway \textit{et al.} Experiment}
The details of the experiment of Broadway \textit{et al.}~\cite{Broadway2018spat} investigating band bending in the commonly used oxygen-terminated diamond are as follows. In that experiment, they performed ODMR spectroscopy on N$V$ centers. They compared the eight resonance frequencies of the N$V^-$ ODMR spectrum to the standard N$V$ spin Hamiltonian including the Zeeman and Stark effects to extract the electric field~\cite{Dolde2011electric,Doherty2012theory}. They found an average electric field in the $z$ direction of $\left<\mathcal{E}_z\right> = 291 \pm 5$~kV~cm$^{-1}$ for the N$_\text{C}$ concentration of $n_{\text{N}_\text{C}} \approx 1.41\times10^{18}$~cm$^{-3}$. Only at this value of $n_{\text{N}_\text{C}}$ was the average electric field not significantly different from the value they obtained in a comparison with N$V$ centers in hydrogen-terminated diamond at the same value for $n_{\text{N}_\text{C}}$. The result suggests that the concentration $n_{\text{N}_\text{C}} \approx 1.41\times10^{18}$~cm$^{-3}$ yields defects placed sufficiently far apart that a typical measured defect will be negligibly influenced by the surface. The implanted ion was $^{15}$N$^+$ at energies ranging from 4 to 20~keV. The ion dose was 10$^{13}$ ions cm$^{-2}$. The nitrogen ions were implanted to form N$V$ centers following a spatial distribution that could be approximated as uniform over the depth range $0-2\left<d\right>$ where $\left<d\right>$ is the average implantation depth. The diamond was electronic grade with an intrinsic substitutional N (N$_\text{C}$) concentration less than $1$~ppb. They modeled the electric field as being induced by surface defects with concentrations as high as 1 nm$^{-2}$, which predicts a maximum electric field value at the surface of $\mathcal{E}_z \approx 1.6$~MV~cm$^{-1}$ with a characteristic decay length of approximately 15~nm. They further argued that a positive space charge density exists near the surface such that only N$V$s deeper than approximately 7~nm for $\left<d\right> = 10$~nm exist in the negative charge state usable for sensing. Therefore, averaging over the N$V^-$ distribution, they estimated a maximum average electric field of $\left<\mathcal{E}_z\right> \approx \int_{7}^{20}1.6~\text{MV~cm}^{-1} e^{-z/15}dz{\big /}\int_{7}^{20}dz \approx 600~\text{kV~cm}^{-1}$ for $\left<d\right> = 10$~nm. By contrast, for $\left<d\right> = 35$~nm, their prediction for the maximum average electric field was $\left<\mathcal{E}_z\right> \approx 200~\text{kV~cm}^{-1}$.

\subsection{Explanation of the Bulk Value of $\left<\mathcal{E}_z\right>$ in the Broadway \textit{et al.} Experiment}
We argue that the Broadway \textit{et al.}~\cite{Broadway2018spat} experiment also captures the band bending due to the built-in electric field between N$_\text{C}^+$ and N$V^-$. For the N$_\text{C}$ concentration of $n_{\text{N}_\text{C}} \approx 1.41\times10^{18}$~cm$^{-3}$, the average electric field measured at the location of N$V^-$ defects should be predominantly due to the built-in electric field between N$_\text{C}^+$ and N$V^-$ rather than due to the surface since hydrogen-terminated and oxygen-terminated samples show little difference between their average electric fields for that concentration of $n_{\text{N}_\text{C}}$~\cite{Broadway2018spat}. Given an implantation dose of 10$^{13}$ ions cm$^{-2}$, the concentration of N$_\text{C}$ is anywhere from 1.29$\times10^{18}$~cm$^{-3}$ to 1.42$\times10^{18}$~cm$^{-3}$ for $\left<d\right> = 35~\text{nm}$, using $\chi$ ranging from $0.1$ to $0.004$ and $n_{\text{N}_\text{C}} = (1-\chi)\cdot\frac{10^{13}~\text{cm}^{-2}}{2\left<d\right>}$~\cite{Broadway2018spat}. Therefore, if the implanted diamond region is partitioned into cubes of equal volume each containing on average a single N$_\text{C}$, the side length of one of these cubes will be anywhere from $l_{\text{N}_\text{C}} = 9.20$~nm to $l_{\text{N}_\text{C}} = 8.89$~nm. The average distance between the N$_\text{C}$ will be an upper bound to the average distance between the N$_\text{C}$ defects and the N$V$ defects since the concentration of N$V$ defects produced by the N$^+$ implantation is approximately 0.4-10\% of the concentration of N$_\text{C}$ defects produced by the implantation~\cite{Broadway2018spat,Pezzagna2010creation}. Without knowing a priori the cutoff distance beyond which charge transfer cannot occur between the species, we assume that charge transfer can occur for any possible separation between the N$V$ and the N$_\text{C}$ within one of the cubes. Thus, averaging over the possible displacements $\mathbf{r}$ between the positions $\mathbf{r}_{\text{N}V}$ and $\mathbf{r}_{\text{N}_\text{C}}$ of the respective N$V^-$ and N$_\text{C}^+$ defects within one of the cubes we have 
\begin{align}
    \label{eq:partials}\left<\mathcal{E}_z\right>_{\left<d\right> = 35~\text{nm}} &= \frac{1}{e}\left<\frac{\partial E_\text{V}}{\partial z}\right>_{\left<d\right> = 35~\text{nm}} \\&\approx \frac{1}{e}\left<\frac{(E_\text{V}(\mathbf{r}_{\text{N}V})-E_\text{V}(\mathbf{r}_{\text{N}_\text{C}}))}{|\mathbf{r}|}\cdot\frac{ z}{|\mathbf{r}|}\right>_{\left<d\right> = 35~\text{nm}}\label{eq:firstexpression}\\&\approx \frac{1}{e}{\bigg(}\int_{x = 0}^{l_{\text{N}_\text{C}}/2}\int_{y=0}^{l_{\text{N}_\text{C}}/2}\int_{z = 0}^{l_{\text{N}_\text{C}}/2}\frac{(E_\text{V}(\mathbf{r}_{\text{N}V})-E_\text{V}(\mathbf{r}_{\text{N}_\text{C}}))}{(x^2+y^2+z^2)^{1/2}}\cdot\frac{z}{(x^2+y^2+z^2)^{1/2}}\text{d}z\text{d}y\text{d}x{\bigg /}\label{eq:firstresult}\\&\quad (l_{\text{N}_\text{C}}^3/2){\bigg)}\nonumber\\
    &\approx0.304\cdot\frac{1}{e}\frac{(E_\text{V}(\mathbf{r}_{\text{N}V})-E_\text{V}(\mathbf{r}_{\text{N}_\text{C}}))}{l_{\text{N}_\text{C}}}\\&\approx 300~\text{kV~cm}^{-1}.
\end{align}

\begin{figure}[ht!] 
\centering
\includegraphics[width=0.6\textwidth]{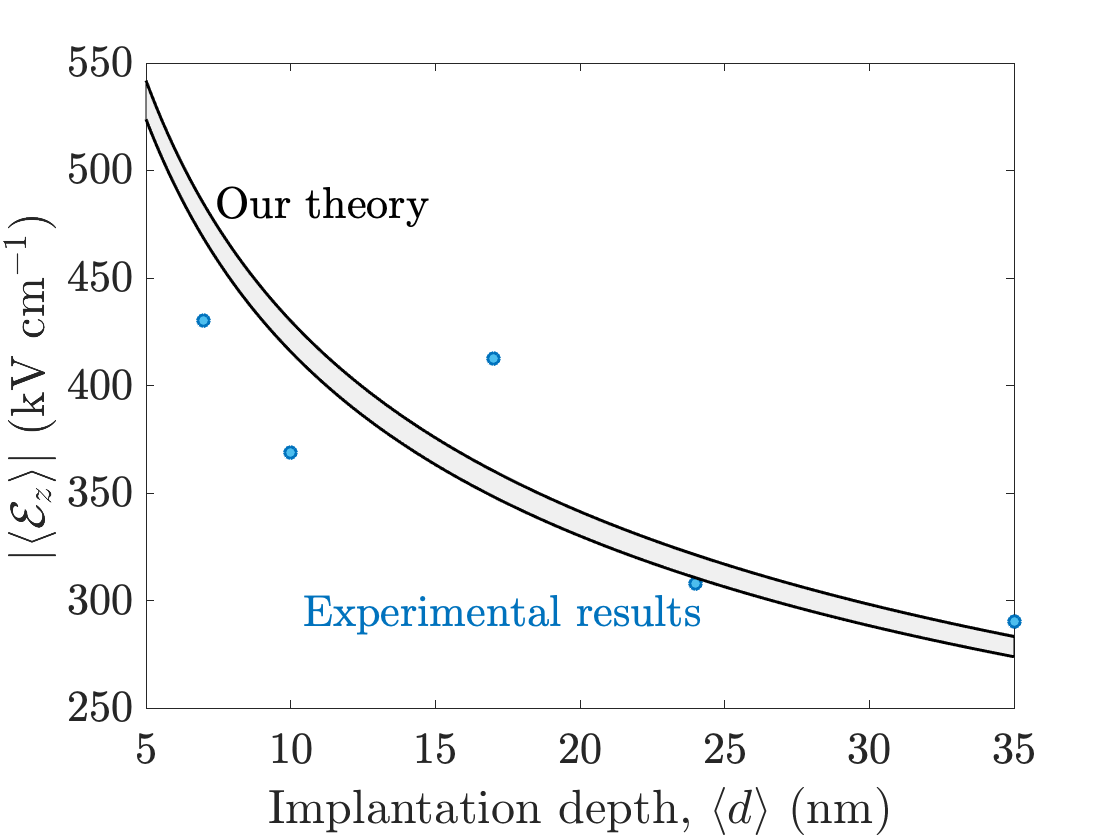}
\caption{Comparison of our theory and the Broadway \textit{et al.}~\cite{Broadway2018spat} experimental results for $\left|\left<\mathcal{E}_z\right>\right|$ for oxygen-terminated diamond. Our theory predicts  $\left|\left<\mathcal{E}_z\right>\right| \approx 4.30~\text{kV~cm}^{-2/3}/\left<d\right>^{1/3}$ for $\chi = 0.004$ (upper solid black curve) and $\left|\left<\mathcal{E}_z\right>\right| \approx 4.16~\text{kV~cm}^{-2/3}/\left<d\right>^{1/3}$ for $\chi = 0.1$ (lower solid black curve), neglecting the effect of the surface. The shaded grey region indicates predictions for the range $\chi=0.1-0.004$. Concentrations for N$_\text{C}$ can be obtained from the abscissae of the plot using $n_{\text{N}_\text{C}} = (1-\chi)\cdot\frac{10^{13}~\text{cm}^{-2}}{2\left<d\right>}$.} 
\label{fig:comparison}
\end{figure}

We note that the factor of 0.304 would change depending on the shape and dimensionality of the supercell containing the two defects. Generally, the integration should proceed over the appropriate Wigner-Seitz cell. As alluded to above, Broadway \textit{et al.}~\cite{Broadway2018spat} provided $\left<\mathcal{E}_z\right>_{\left<d\right> = 35~\text{nm}} = 291\pm5$~kV~cm$^{-1}$, which is in good agreement with our calculated value. Keeping additional significant digits in our calculation, the difference between the absolute values of the results is anywhere from less than 3\% for $\chi = 0.004$ to less than 6\% for $\chi = 0.1$. The high nitrogen fluence used in the Broadway \textit{et al.} experiment suggests $\chi \approx 0.01$~\cite{Broadway2018spat}, corresponding to a difference between our calculated result and experiment of less than 3\%. By contrast, our application of the same formalism for $\left<d\right> = 7~\text{nm}$ yields $\left<\mathcal{E}_z\right>_{\left<d\right> = 7~\text{nm}} \approx 500~\text{kV~cm}^{-1}$ in comparison with $\left<\mathcal{E}_z\right>_{\left<d\right> = 7~\text{nm}} = 432 \pm 10~\text{kV~cm}^{-1}$ from experiment~\cite{Broadway2018spat}. Keeping additional significant digits, the difference between the absolute values of the results is anywhere from approximately 12\% for $\chi = 0.004$ to approximately 8\% for $\chi = 0.1$. A complete comparison of our theory and the Broadway \textit{et al.}~\cite{Broadway2018spat} experimental results for $\left|\left<\mathcal{E}_z\right>\right|$ for oxygen-terminated diamond is depicted in Fig. \ref{fig:comparison}. We find that for the smaller N$_\text{C}$ concentrations, $n_{\text{N}_\text{C}} \lesssim 2.1\times10^{18}$~cm$^{-3}$, the band bending due to the built-in field between N$_\text{C}^+$ and N$V^-$ appears to determine the average electric field measured at the location of N$V^-$ centers. 

We now turn to a discussion of our derivation. In moving from Eq. (\ref{eq:firstexpression}) to Eq. (\ref{eq:firstresult}), we have used the fact that there are four possible N$V^-$ orientations in the diamond crystal. Only one of these contributes to an appreciable measured field in the $z$ direction since $\mathcal{E}_x$ and $\mathcal{E}_y$ were set to zero in the reference frame of each N$V^-$ in fitting the measured spectra in the work of Broadway \textit{et al.}~\cite{Broadway2018spat}. To arrive at our results, we have also used  $\left(\epsilon^{\rm{N}_\text{C}}(0/+)-\epsilon^{{\rm N}V}(0/-)\right) \approx 0.8$~eV from our calculations. Finally, we have used the fact that the Broadway \textit{et al.}~\cite{Broadway2018spat} measurements were insensitive to the sign of the electric field so that the contribution from integrating over negative $z$ does not cancel the contribution from integrating over positive $z$ and we need only consider the absolute value of our calculated result. We also note that our explanation requires no fitted parameters as long as $n_{\text{N}_\text{C}}$ is known and that our results would be generally applicable to defects in any wide-bandgap semiconductor, such as defects in the various widely studied polytypes of SiC and other defects in diamond~\cite{Kraus2014room,Castelletto2014a,Bockstedte2004ab,Koehl2011room,Riedel2012resonant,Kimoto2014fundamentals,Wang2017efficient,Wang2017scalable,Fuchs2015engineering,Kuate2018energetics,Gadalla2021enhanced,Kuate2019parallel,Kuate2021calculating,Nagy2018quantum,Widmann2015coherent,Lohrmann2017a,Bracher2017selective,Falk2013polytype,Soykal2016silicon,Soykal2017quantum,Weber2010quantum,Kraus2017three,Gali2011time,Awschalom2018quantum,Wolfowicz2021quantum,Whiteley2019spin}.

We now comment on the validity of the neglect of second-nearest neighbors. We first note that due to charge conservation, most of the N$_\text{C}$ defects that are nearest neighbors to the N$V$ will be neutral and will not contribute to the field at the N$V$. Furthermore, the charged N$V$-N$_\text{C}$ defect pairs surrounding the N$V$-N$_\text{C}$ pair of interest will be randomly positioned. As we have shown in the Section \ref{sec:motivation}, the timescale for the equilibration of $E_\text{F}$ is not sufficient for equilibration between more than on average a single pair of defects. Therefore, the charge states of other N$V$-N$_\text{C}$ pairs will be uncorrelated with the charge states of the pair of interest so that simply measuring the system at different points as it evolves forward in time will result in fluctuations of the field in time. These fluctuations will cancel averaged over time for defect pairs that are sufficiently deep in the bulk and yield the value obtained by integrating Eq. (\ref{eq:Efield}) over a supercell containing on average a single defect pair normalized by volume. Thus, our model applies to a system containing arbitrary concentrations $n_{\text{N}V}$ and $n_{\text{N}_\text{C}}$, as long as these are in the dilute limit and as long as the N$V$ is measured sufficiently deep in the bulk. We further clarify that Eq. (\ref{eq:eqconc}) is a local statement that is automatically satisfied if $1/\bar{\Gamma}$ is much longer than the timescale of experimental measurements. The equation does not require that there be one N$V$ for every N$_\text{C}$ in the entire sample. The existence of a positive space charge density near the surface in the experiment of Broadway \textit{et al.}~\cite{Broadway2018spat} could explain the discrepancy between our theory and experiment at the larger N$_\text{C}$ concentrations. We also note that in the moment of measurement the wavefunction of the electron collapses so that the value obtained by integrating Eq. (\ref{eq:Efield}) over a supercell containing on average a single defect pair normalized by volume does indeed represents the value of the field that will be measured at the N$V$.

We now turn to a discussion of the implications of our results for the phenomenon of spectral diffusion. Formally, we can consider neighboring N$V$-N$_\text{C}$ pairs in an infinite crystal so that we have
\begin{align}
    \label{eq:partialssum}\left<\mathcal{E}_z\right> &\approx \frac{1}{e}{\bigg(}\sum_{n_x,n_y,n_z\in\mathbb{Z}}\int_{x = -l_{\text{N}_\text{C}}/2}^{l_{\text{N}_\text{C}}/2}\int_{y=-l_{\text{N}_\text{C}}/2}^{l_{\text{N}_\text{C}}/2}\int_{z = --l_{\text{N}_\text{C}}/2}^{l_{\text{N}_\text{C}}/2}\frac{(E_\text{V}(\mathbf{r}_{\text{N}V})-E_\text{V}(\mathbf{r}_{\text{N}_\text{C}}))}{((x+n_xl_{\text{N}V})^2+(y+n_yl_{\text{N}V})^2+(z+n_zl_{\text{N}V})^2)^{1/2}}\times\\&\frac{(z+n_zl_{\text{N}V})}{((x+n_xl_{\text{N}V})^2+(y+n_yl_{\text{N}V})^2+(z+n_zl_{\text{N}V})^2)^{1/2}}\text{d}z\text{d}y\text{d}x{\bigg /} (l_{\text{N}_\text{C}}^3){\bigg)}.\nonumber
\end{align}
Above, $l_{\text{N}V} = n_{\text{N}V}^{-1/3}$. By symmetry, $\left<\mathcal{E}_z\right>$ is zero. If, however, we partition the infinite crystal into two half-spaces, $n_z \geq 0$ and $n_z < 0$, for any N$V$ in either half-space the infinite sum diverges. Therefore, a N$V$ belonging to a defect pair that is sufficiently deep in the bulk will show negligible spectral diffusion, which is not true for a N$V$ belonging to a defect pair near the surface. We therefore estimate the maximum spectral diffusion for a N$V$ at the surface in the experiment of Ruf \textit{et al.}~\cite{Ruf2019optically}. In that experiment, they used a sample that was 2 mm $\times$ 2 mm $\times$ 50 $\mu$m. The highest estimate for $n_{\text{N}V}$ was $n_{\text{N}V} = 0.1$ $\mu$m$^{-3}$ corresponding to $l_{\text{N}V} = 2$ $\mu$m. The highest estimate for $n_{\text{N}_\text{C}}$ was $n_{\text{N}_\text{C}} = 885$ $\mu$m$^{-3}$ corresponding to $l_{\text{N}_\text{C}} = 0.104$ $\mu$m. Thus, to calculate the electric field at a surface N$V$, $n_z$ ranges from 0 to 23 and $n_x$ and $n_y$ both range from -464 to 464. Except for $(n_x,n_y,n_z) = (0,0,0)$, we drop the integral over $x$, $y$, and $z$ (setting the variables $x$, $y$, and $z$ to zero) and drop the normalizing volume factor of $l_{\text{N}_\text{C}}^3$ in evaluating the contributions. We obtain a field of approximately 24,000~kV cm$^{-1}$, corresponding to a spectral diffusion of approximately 400~MHz~\cite{Dolde2011electric,Broadway2018spat}. The experiment of Ruf \textit{et al.}~\cite{Ruf2019optically} produced a confidence interval for spectral diffusion of 189$\pm$117 MHz whose upper end is in reasonable agreement with our maximum estimate. We would also like to emphasize the utility of our formalism for estimating average defect concentrations or average distances between defects.

\section{CONCLUSION \label{sec:conc}}
In conclusion, based purely on $ab~initio$ calculations, we have succeeded in providing an explanation for the value of the average electric field of 291$\pm$5~kV~cm$^{-1}$ for the N$_\text{C}$ concentration of $n_{\text{N}_\text{C}} \approx 1.41\times10^{18}$~cm$^{-3}$ for the commonly used oxygen-terminated diamond. Such a result would be useful for predicting the functioning of semiconductor devices as rectifiers and switching devices, where the built-in defect-induced fields would lead to losses. Our results could also be useful for predicting and correcting the spectral diffusion of the optical frequencies of the solid-state single-photon sources used for applications in quantum information and computation. Furthermore, our formalism for thermally driven charge transfer could aid in investigations of charge dynamics~\cite{Bluvstein2019identifying,Yuan2020charge,Capelli2022proximal}.  

\section*{ ACKNOWLEDGMENTS:}
R.K.D. gratefully acknowledges financial support from the Princeton Presidential Postdoctoral Research Fellowship and from the National Academies of Science, Engineering, and Medicine Ford Foundation Postdoctoral Fellowship program. We also acknowledge support by the STC Center for Integrated Quantum Materials, NSF Grant No. DMR-1231319.  We also thank the referees for their many critical and helpful suggestions which have been instrumental in improving the clarity of our paper.

\bibliography{refs_NV}
\end{document}